\documentclass[aps,pra,reprint,superscriptaddress,amsthm,amsmath,amssymb,showpacs]{revtex4-1}

\newtheorem{thm}{Theorem}

\begin{document}


\title{One parameter family of indecomposable optimal entanglement witnesses arising from generalized Choi maps}

\author{Kil-Chan Ha}
\affiliation{Faculty of Mathematics and Statistics, Sejong University, Seoul 143-747, Korea}

\author{Seung-Hyeok Kye}
\affiliation{Department of Mathematics, Seoul National University, Seoul 151-742, Korea}
\date{\today}

\begin{abstract}
In the recent paper [Chru\'{s}ci\'{n}ski and Wudarski, arXiv:1105.4821],
it was conjectured that the entanglement witnesses
arising from some generalized Choi maps are optimal.
We show that this conjecture is true. Furthermore, we show that they provide
a one parameter family of indecomposable optimal entanglement witnesses.
\end{abstract}


\maketitle

Quantum entanglement is a basic resource in quantum information processing and
communication  \cite{nielsen}. Therefore, so much effort naturally
has been put into developing theoretical and experimental methods of entanglement detection.
Among them, one of the most general approach is based on the
notion of entanglement witness \cite{horodecki96,terhal00}.
Recall that an observable $W=W^{\dagger}$ is said to be an entanglement
witness (EW) if $\text{tr}(W\sigma) \ge 0$ for all separable states $\sigma$,
and there exists an entangled
state $\rho$ for which $\text{tr}(W\rho)<0$. In this case we say that
$W$ detects $\rho$.
Following \cite{lkch00}, an EW $W$ is said to be optimal if the set
of entanglement detected by $W$ is maximal with respect to the set inclusion.

Note that the Choi-Jamio\l wski isomorphism \cite{choi75,jamio72}
gives rise to an entanglement witness
$$
W_{\Lambda} =(\openone \otimes \Lambda)P^+,
$$
which is acting on $M_M\otimes M_N$,
for every positive linear map $\Lambda\,:\,M_M\to M_N$ which is not completely positive,
where $M_K$ denotes the $C^*$-algebra of
all $K\times K$ matrices over the complex field $\mathbb C$, and $P^+$ denotes the projector
onto the maximally entangled state in $\mathbb C^M\otimes \mathbb C^M$.
It is well known that decomposable positive maps give decomposable
entanglement witnesses which take the general form
$W=P+Q^{\Gamma}$,
where $P,\, Q \ge 0$ and $\Gamma$ refers to partial transposition with
respect to the second subsystem, that is, $Q^{\Gamma}=(\openone\otimes T)Q$. If a
given witness can not be written in this form, we call it indecomposable.
Of course, indecomposable EWs arise from indecomposable positive maps \cite{lkch00,lkhc01,hakye}

Note that an EW is indecomposable if and only if it detects entangled states
with positive partial transposes \cite{lkch00}. Thus, so far indecomposable EWs are concerned, it is natural
to consider the optimality by requiring the witness to be finer with respect to entangled states with positive partial
transposes only. This kinds of witness is said to be an indecomposable optimal EW.
It is known \cite{lkch00,kabl} that
$W$ is an indecomposable optimal entanglement
witness if and only if both $W$ and $W^{\Gamma}$ are optimal entanglement
witnesses.

Typical examples of indecomposable optimal EW come from indecomposable positive linear maps
which generate an extremal ray of the convex cone consisting of all positive linear maps.
The Choi map is an example of this kind \cite{choi75,choilam,kabl}.
Variations of the Choi map given by the second author \cite{blau} also give rise to
such maps. Some of them, parameterized by three real variables, were shown to be
extremal in \cite{osaka}. See the recent paper \cite{arv} for related topics.
Although there are some examples of optimal EW \cite{kabl,cps,cp,cp11,cw},
to the best of the author's knowledge, only known examples of indecomposable optimal EWs are ones which come from
extremal indecomposable positive linear maps.

We consider another variations of the Choi map given in \cite{ckl}. For nonnegative real numbers $a,b,c$,
we define the linear map $\Phi[a,b,c]:M_3\to M_3$ by
\begin{widetext}
\[
\Phi[a,b,c](X)=\\ \dfrac{1}{a+b+c}\begin{pmatrix}
ax_{11}+bx_{22}+cx_{33} & -x_{12} & -x_{13} \\
-x_{21} & cx_{11}+ax_{22}+bx_{33} & -x_{23} \\
-x_{31} & -x_{32} & bx_{11}+cx_{22}+ax_{33}
\end{pmatrix},
\]
\end{widetext}
for $X=[x_{ij}] \in M_3$.
It was shown that $\Phi[a,b,c]$ is positive if and only if
$$
a+b+c\ge 2,\qquad 0\le a\le 2\implies bc\ge (1-a)^2.
$$
In the case of $a=1$, the maps $\Phi[1,1,0]$ and $\Phi[1,0,1]$ reproduce the Choi map and its dual,
respectively \cite{choi75}.
The matrix representation of EW $W[a,b,c]:=W_{\Phi[a,b,c]}$ is given by
\[
W[a,b,c]=\dfrac  16 \left(\begin{array}{ccc|ccc|ccc}
a & \cdot & \cdot &\cdot & -1 & \cdot &\cdot & \cdot & -1 \\
\cdot & b & \cdot &\cdot & \cdot & \cdot &\cdot & \cdot & \cdot \\
\cdot & \cdot & c &\cdot & \cdot & \cdot &\cdot & \cdot & \cdot \\\hline
\cdot & \cdot & \cdot &c & \cdot & \cdot  &\cdot & \cdot & \cdot \\
-1 & \cdot & \cdot &\cdot & a & \cdot &\cdot & \cdot & -1 \\
\cdot & \cdot & \cdot &\cdot & \cdot & b &\cdot & \cdot & \cdot \\\hline
\cdot & \cdot & \cdot &\cdot & \cdot & \cdot &b & \cdot & \cdot \\
\cdot & \cdot & \cdot &\cdot & \cdot & \cdot &\cdot & c & \cdot \\
-1 & \cdot & \cdot &\cdot & -1 & \cdot &\cdot & \cdot & a
\end{array}
\right),
\]
where we replaced zeros by dots.

Recently, Chru\'{s}ci\'{n}ski and Wudarski \cite{cw} analyzed the following case:
\begin{equation}\label{cond}
0\le a\le 1,\quad a+b+c=2,\quad bc=(1-a)^2,
\end{equation}
and parameterize them by
\begin{align*}
a(\alpha)&=\dfrac{2}{3}(1+\cos(\alpha)),\\
b(\alpha)&=\dfrac{2}{3}\left(1-\dfrac 12 \cos(\alpha)-\dfrac{\sqrt3}2\sin(\alpha)\right),\\
c(\alpha)&=\dfrac{2}{3}\left(1-\dfrac 12 \cos(\alpha)+\dfrac{\sqrt3}2\sin(\alpha)\right),
\end{align*}
for ${\pi}/3\le \alpha \le {5\pi}/3$.
They conjectured that the entanglement witnesses $W[\alpha]=W[a(\alpha), b(\alpha), c(\alpha)]$
arising from them are optimal for every $\alpha$ with ${\pi}/3\le \alpha \le {5\pi}/3$.

The purpose of this Brief Report is to show that this conjecture is true.
Furthermore, we show that $W[\alpha]$ is an indecomposable optimal EW for
each $\pi/3 < \alpha <\pi$ and $\pi <\alpha <5\pi/3$.
Equivalently, we show the following:

\begin{thm}\label{conj}
If $a,\,b,\,c$ are nonnegative numbers satisfying the conditions~\eqref{cond}
then entanglement witnesses $W[a,b,c]$ are optimal.
Furthermore, $W[a,b,c]$ is an indecomposable optimal entanglement witness,
whenever $0<a<1$.
\end{thm}

For an EW $W$, define
$$
\mathcal P_{W}=\{\psi\otimes \phi \in \mathcal H_A\otimes \mathcal H_B\,:\,
\langle \psi\otimes \phi |W|\psi \otimes \phi\rangle=0\}.
$$
It is known \cite{lkch00} that
if the set $\mathcal P_W$ spans the entire Hilbert space $\mathcal H_A\otimes \mathcal H_B$, then $W$ is an optimal EW.


{\bf Proof:}
It suffices to consider the case $0\le a<1$.  Put $t=c/(1-a)$ for $a,\,c$ satisfying the
condition \eqref{cond}. Then $t$ is a positive number and satisfies the following conditions:
\begin{equation}\label{xxx}
a+bt =1,\qquad
c+at =t.
\end{equation}
For each $k=1,2,\cdots,9$, we define vectors $|\psi_k\rangle,\,|\phi_k\rangle$ in $\mathbb C^3$ as follows:
\[
\begin{aligned}
|\psi_1\rangle &=|0\rangle +|1\rangle +|2\rangle , \\
|\psi_2\rangle &=|0\rangle -|1\rangle +|2\rangle , \\
|\psi_3\rangle &=|0\rangle +i|1\rangle -i|2\rangle , \\
|\psi_4\rangle &=\sqrt{t}|1\rangle +|2\rangle , \\
|\psi_5\rangle &=\sqrt{t}|1\rangle +i|2\rangle , \\
|\psi_6\rangle &=|0\rangle +\sqrt{t}|2\rangle , \\
|\psi_7\rangle &=i|0\rangle +\sqrt{t}|2\rangle , \\
|\psi_8\rangle &=\sqrt{t}|0\rangle +|1\rangle ,  \\
|\psi_9\rangle &=\sqrt{t}|0\rangle +i|1\rangle ,\\
\end{aligned}
\quad
\begin{aligned}
|\phi_1\rangle &=|0\rangle +|1\rangle +|2\rangle , \\
|\phi_2\rangle &=|0\rangle -|1\rangle +|2\rangle ,\\
|\phi_3\rangle &=|0\rangle -i|1\rangle +i|2\rangle ,\\
|\phi_4\rangle &=\sqrt{t}|1\rangle +t|2\rangle ,\\
|\phi_5\rangle &=\sqrt{t}|1\rangle -t i|2\rangle ,\\
|\phi_6\rangle &=t|0\rangle +\sqrt{t}|2\rangle ,\\
|\phi_7\rangle &=-t i|0\rangle +\sqrt{t}|2\rangle ,\\
|\phi_8\rangle &=\sqrt{t}|0\rangle +t|1\rangle  ,\\
|\phi_9\rangle &=\sqrt{t}|0\rangle -t i|1\rangle .\\
\end{aligned}
\]

For $k=1,2,3$, it is easy to see that
\begin{align*}
&\langle \psi_k\otimes \phi_k | W[a,b,c]|\psi_k\otimes \phi_k\rangle \\
=&\langle \psi_k\otimes \phi^*_k |( W[a,b,c])^{\Gamma}|\psi_k\otimes \phi^*_k\rangle\\
=& 3(a+b+c-6)=0.
\end{align*}
Using the condition (\ref{xxx}), we also have
\begin{align*}
  &\langle \psi_k \otimes \phi_k |W[a,b,c]|\psi_k\otimes \phi_k\rangle\\
=&\langle \psi_k\otimes \phi^*_k |( W[a,b,c])^{\Gamma}|\psi_k\otimes \phi^*_k\rangle\\
=&bt^3+2(a-1)t^2+ct\\
=&(at+c)t+(a+bt)t^2-2t^2,\\
=&t^2+t^2-2t^2=0
\end{align*}
for $k=4,5,\cdots,9$. So all
the vectors $\psi_k \otimes \phi_k$ ($\psi_k \otimes \phi^*_k$ respectively)
belong to the set $\mathcal P_{W[a,b,c]}$ ($\mathcal P_{(W[a,b,c])^{\Gamma}}$ respectively).

Now, we show that  the set
\begin{equation}\label{prod-1}
\{\psi_k\otimes \phi_k:1\le k\le 9\}
\end{equation}
spans the entire space $\mathbb C^3\otimes \mathbb C^3$ for $0\le a<1$, and the set
\begin{equation}\label{prod-2}
\{\psi_k\otimes \phi^*_k:1\le k\le 9\}
\end{equation}
spans the entire space for $0<a<1$.
Let $M$ ($M'$ respectively) be
the $9\times 9$ matrix whose $k$-th column
is $|\psi_k \otimes \phi_k\rangle$ ($|\psi_k \otimes \phi^*_k\rangle$ respectively).
Then the determinant of $M$ and $M'$ are given by
\begin{align*}
|M|=&8t^4(t^2-1)\sqrt{t}(2t-\sqrt{t}+2)\\
        &\ -8t^5(1+t)(t-4\sqrt{t}+1)i,\\
|M'|=&-8t^4\sqrt{t}(t-1)^3-8t^4\sqrt{t}(t-1)^3i.
\end{align*}
The only positive solution of the equation $\text{Re}(|M|)=0$ is $t=1$,
where $\text{Re}(z)$ denotes the real part of complex number $z$.
But for this $t=1$, we have $|M|\neq 0$. Therefore we can conclude
that $|M|\neq 0 $ for positive number $t$.
Consequently, the set (\ref{prod-1}) spans the entire space, and so $W[a,b,c]$ is an optimal EW.

On the other hand, the only positive solution of the equation $|M'|= 0$ is $t=1$. Since $t=1$
is equivalent to $(a,b,c)=(0,1,1)$, the set (\ref{prod-2}) spans
the entire space except for the case $(a,b,c)=(0,1,1)$.
Therefore, we conclude that $W[a,b,c]$ is an indecomposable optimal EW
except for the case of $a=0$.

When $a=0$, $\Phi[0,1,1]$ is a decomposable positive map, and the optimality of
$W[0,1,1]=W[\pi]$ is known in \cite{cw}. This completes the proof of Theorem~\ref{conj}.

In conclusion, we proved that the conjecture posed by Chru\'{s}ci\'{n}ski and Wudarski is true.
Moreover, we provided a one parameter family of indecomposable optimal EWs.
In \cite{cw}, Chru\'{s}ci\'{n}ski and Wudarski  showed that  the structural physical
approximation  defines a separable state for a large class of $W[\alpha]$.
So our result supports the another conjecture \cite{kabl} that the SPA to an optimal entanglement witness defines a separable state.
Finally, it woul be interesting to know if the map $\Phi[a,b,c]$ generates an extremal ray of the cone of all positive linear maps
whenever the condition (\ref{cond}) holds.

\begin{acknowledgments}
The first (respectively second) author was partially supported by Basic Science Research Program through the
National Research Foundation of Korea(NRF) funded by the Ministry of Education, Science
and Technology (NRFK 2011-0006561) (respectively (NRFK 2011-0001250)
\end{acknowledgments}

\end{document}